 \documentclass[final,1p,times,twocolumn,authoryear]{elsarticle}

\usepackage{amssymb}

\usepackage{textcomp}
\DeclareTextSymbolDefault{\degreesymbol}{TS1} 
\DeclareTextSymbol{\degreesymbol}{TS1}{176} 
\DeclareRobustCommand{\textdegree}{\ifmmode\mbox{\degreesymbol}\else\degreesymbol\fi}

\usepackage{graphicx}	
\usepackage{amsmath}	
\usepackage{amssymb}	
\usepackage{rotating}
\usepackage{lscape}
\usepackage{comment}
\usepackage[utf8]{inputenc}
\usepackage[dvipsnames]{xcolor}
\usepackage{siunitx}
\usepackage{verbatim}
\usepackage[english]{babel}
\usepackage[T1]{fontenc}
\usepackage{ae,aecompl}
\usepackage{rotating}
\usepackage{textcomp}
\usepackage{lipsum}
\usepackage{url}
\usepackage[normalem]{ulem}
\usepackage{soul}
\usepackage[switch]{lineno}

\begin{document}
\newcommand{\glm}[1]{\textcolor{ForestGreen}{#1}}



\title{4FGLzoo. Classifying {\it Fermi}-LAT uncertain gamma-ray sources by machine learning analysis.}

 \author[label1,label2]{Graziano Chiaro}
 \address[label1]{Istituto di Fisica Cosmica e Astrofisica Spaziale  IASF, INAF Via A. Corti 12,  20133 Milano  IT}  
 \address[label2]{Consorzio Interuniversitario per la Fisica Spaziale  CIFS ,Via Pietro Giuria, 1, 10125 Torino IT}  
 \address[label3]{Lab. de Instrumenta\c{c}\~ao e F\'{i}sica Experimental de Part\'{i}culas LIP,Av. Prof. Gama Pinto 2, 1649-003 Lisboa, PT}
\author[label1]{Milos Kovacevic}
\author[label3]{Giovanni La Mura}

\begin{abstract}
Since 2008 August the Fermi Large Area Telescope (LAT) has provided a continuous coverage of the gamma-ray sky yielding more than 5000 $\gamma$-ray sources, but 54$\%$ of the detected sources remain with no certain or unknown association with a low energy counterpart. Rigorous determination of class type for a $\gamma$-ray source requires the optical spectrum of the correct counterpart but optical observations are demanding and time-consuming, then machine learning techniques can be a powerful alternative for screening and ranking. We use machine learning techniques  to select blazar candidates among uncertain sources characterized by $\gamma$-ray properties very similar to those of Active Galactic Nuclei. Consequently, the percentage of sources of uncertain type drops from 54$\%$ to less than 12$\%$ predicting a new zoo for the Fermi $\gamma$-ray sources. The result of this study opens up new considerations on the population of the gamma energy sky, and it will facilitate the planning of significant samples for rigorous analysis and multi-wavelength observational campaigns.
\end{abstract}





\maketitle

\section{Introduction}
\label{<1>}
Since the beginning of its mission, the Fermi Large Area Telescope \citep{lat} detected more than 5000 $\gamma$-ray sources in the 100 MeV - 300 GeV energy range. 
Referring to 10 years of monitoring \citep{dr2}
in the fourth Fermi Large Area Telescope catalog (4FGL) the two largest classes of sources are Active Galactic Nuclei (AGN) and pulsars (PRS).Approximately 95$\%$ of AGNs can be classified as blazar objects subdivided in BL Lacertae (BLL) and Flat Spectrum Radio Quasars (FSRQ) or uncertain sources positionally coincident with an object showing distinctive broad-band blazar characteristics (BCU), but lacking reliable optical spectrum measurements. 
BCUs represent 43$\%$ of the blazar class. Also, 28$\%$ of the 
Fermi-LAT sources have not even a tentative association with a likely $\gamma$-ray emitting object and are referred to as Unassociated Gamma-ray Sources (UGS). As a result, the nature of $\sim$ 54$\%$ of the $\gamma$-ray sources is not yet completely known.\\ However, since blazars are the most numerous $\gamma$-ray source class, we expect that a large fraction of the uncertain or unassociated sources might belong to one of the blazar subclasses, BLL or FSRQ.\\
FSRQs show strong, broad emission lines at optical wavelengths, while BLL spectra are more challenging since they show at most weak emission lines, but they are often  featureless or can display absorption features\citep{Abdo01}
Unfortunately optical observations are demanding and time-consuming. Machine learning techniques (ML) can be a powerful
tool for screening and ranking gamma sources according to their classification.
We tested the power of ML methods by applying a ML algorithm to the first release of the Fermi Large Area Telescope Fourth Source Catalog \citep{4fgl}.
This paper is a continuation of a series of 
studies that use machine learning to classify Fermi Large Area Telescope gamma-ray sources that are likely active galactic nuclei \citep{bflap, 3zoo}.\\
In this study, we use an improved release of an original Artificial Neural Network (ANN) algorithm 
first applied in \citet{bflap} and based on the variation of the Empirical Cumulative Distribution Function \citep[ECDF,][]{kol} extracted from the $\gamma$-ray light curves of the sources. A detailed description of the algorithm that we use and of its performance dealing with the BCU population in the 3FGL data set is described in \citet{milos}
 This paper is organized as follows: in Sect.~\ref{<2>} we provide a brief description of the machine learning technique. In that section, we also 
 discuss the result of the ANN analysis. In Sect.~\ref{<3>} we validate the predictions by optical spectral observations of a number of targets, finally we discuss the 4FGL zoo in Sect.~\ref{<4>}.

\section{Machine learning technique}
\label{<2>}
In previous studies, machine learning techniques, mainly ANN, have been applied to classifying uncertain $\gamma$-ray sources \citep[see, e.g.,][and others]{ack, lee, has, doe, bflap, mira, saz, lef, 3zoo}\\
The basic building block of an ANN algorithm is the neuron. 
Information is passed as inputs to the {\it{neuron}}, which produces an output \citep{gish, ric}. The output is typically determined as a 
mathematical function of the inputs and  can be interpreted as a Bayesian a posteriori probability that models the likelihood of classification 
based on input parameters. The power of a ANN algorithm  comes from assembling many neurons into a network. The network can model very complex behavior from input to output.\\
Here we apply the same ANN algorithm described in \citet{milos} to the data set available at the time of the 4FGL catalog publication \citep{4fgl} which covered 8 years of observations. The improvement over the previous study is due to the use of a longer observing period, including the first eight years of operations, to better statistics on the measurements of the source parameters, and a wider database of multi-wavelength information. The only relevant change in the process consists in the use of the 7 energy bands for the measurements of energy-dependent properties in 4FGL (0.05-0.1, 0.1-0.3, 0.3-1, 1-3, 3-10, 10-30, 30-300 GeV), with respect to the 5 bands used in 3FGL.

\subsection{Analysis of $\gamma$-ray sources}
In \citet{bcu}, the authors analyzed 1329 
BCUs and the algorithm selected 801 BLL candidates, 406 FSRQ candidates, while 122 sources remained with uncertain classification.
Zhu et al. (2020) used a combination of Random Forest (RF) and ANN classification approaches to study 1336 UGS sources. As result 583 sources were classified as AGN-like, 115 as pulsars, and 154 as sources belonging to other classes of known gamma-ray emitters while 484 sources remained of uncertain classification.\\ 
Since the algorithm used in our study is specific for blazar classes selection we further extended the classification of these UGS sources, by providing a blazar classification likelihood for the ones identified as AGN candidates.\\
Machine learning techniques always consider the algorithm reliable efficiency to produce classification likelihoods if it achieves a precision of 90$\%$ {\footnote{ Precision is a threshold based on the optimization of the positive association rate defined as the fraction of true positives over the objects classified as positive, of $\sim$ 90$\%$.}
In Fig.~\ref{ucs3} the BL Lac likelihood L$_{B}$ for each of the 583 sources in our sample is shown. Each source is presented by a green dot. On the left-hand vertical axis, the final value of L$_{B}$ is an average of 300 L$_{B}$ obtained by selecting different training and testing samples. On the right-hand side of the plot are the cumulative values.
The cumulative value for each source is obtained by summing all the  L$_{B}$ values (from the left-hand vertical axis) above a given  L$_{B}$ for that source and dividing by the number of sources whose  L$_{B}$ is above the given  L$_{B}$. 
If we condiser all the sources with a cumulative value above 0.9, we expect 90$\%$ of them to be genuine BLLac with a 10$\%$ contamination of genuine FSRQ. The cumulative values for FSRQs can be  obtained in a similiar way.\\
The light green area corresponds to 1$\sigma$ error due to differences in train-test sample selections. With 90$\%$ precision the algorithm identified {\bf{294}} BLL candidates ( L$_{B}$ $\geq$ 0.639 ) and {\bf{164}} FSRQs ( L$_{B}$ $\leq$ 0.254) while {\bf{125}} sources remained of uncertain classification. Fig.~\ref{ucsh} presents the histogram of classification likelihoods. Two distinct and opposite peaks for BL Lac (blue) and FSRQ (red), the former at L$_{BLL}$ $\sim$1, while the latter at L$_{B}$ $\sim$0 are visible.\\

\begin{figure}
\includegraphics[width=.8\textwidth]{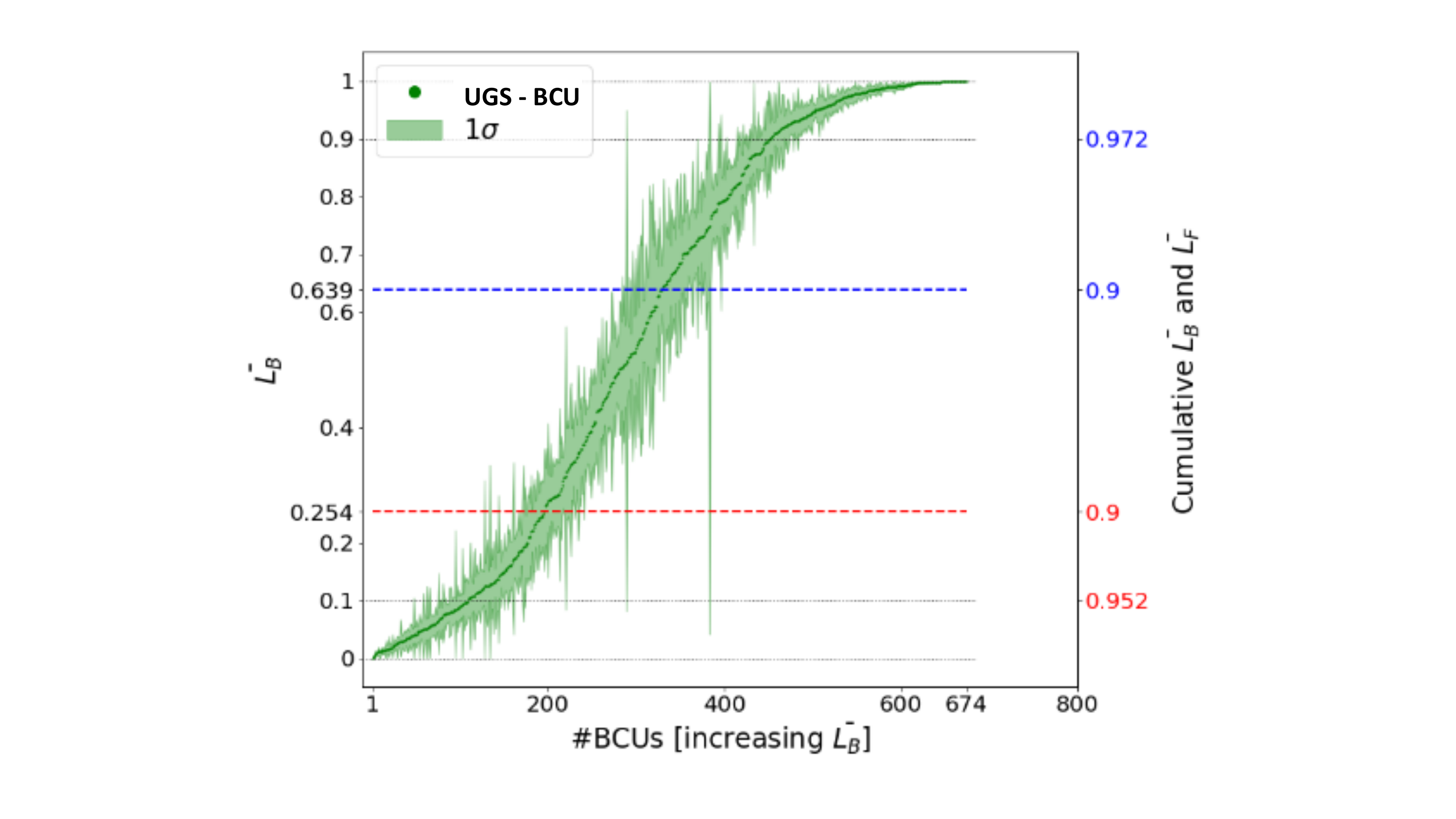}
\caption{BL Lac probability $L_B$ $(= 1 - L_F)$      of 583 UGS$_{BCU}$. Each source is presented by a green dot. The right-hand vertical axis shows the corresponding cumulative L$_{B}$ (blue) and L$_{F}$ (red). The light green area corresponds to 1$\sigma$ error due to
differences in train-test sample selections.\label{ucs3}}
\end{figure}

\begin{figure}
\includegraphics[width=.8\textwidth]{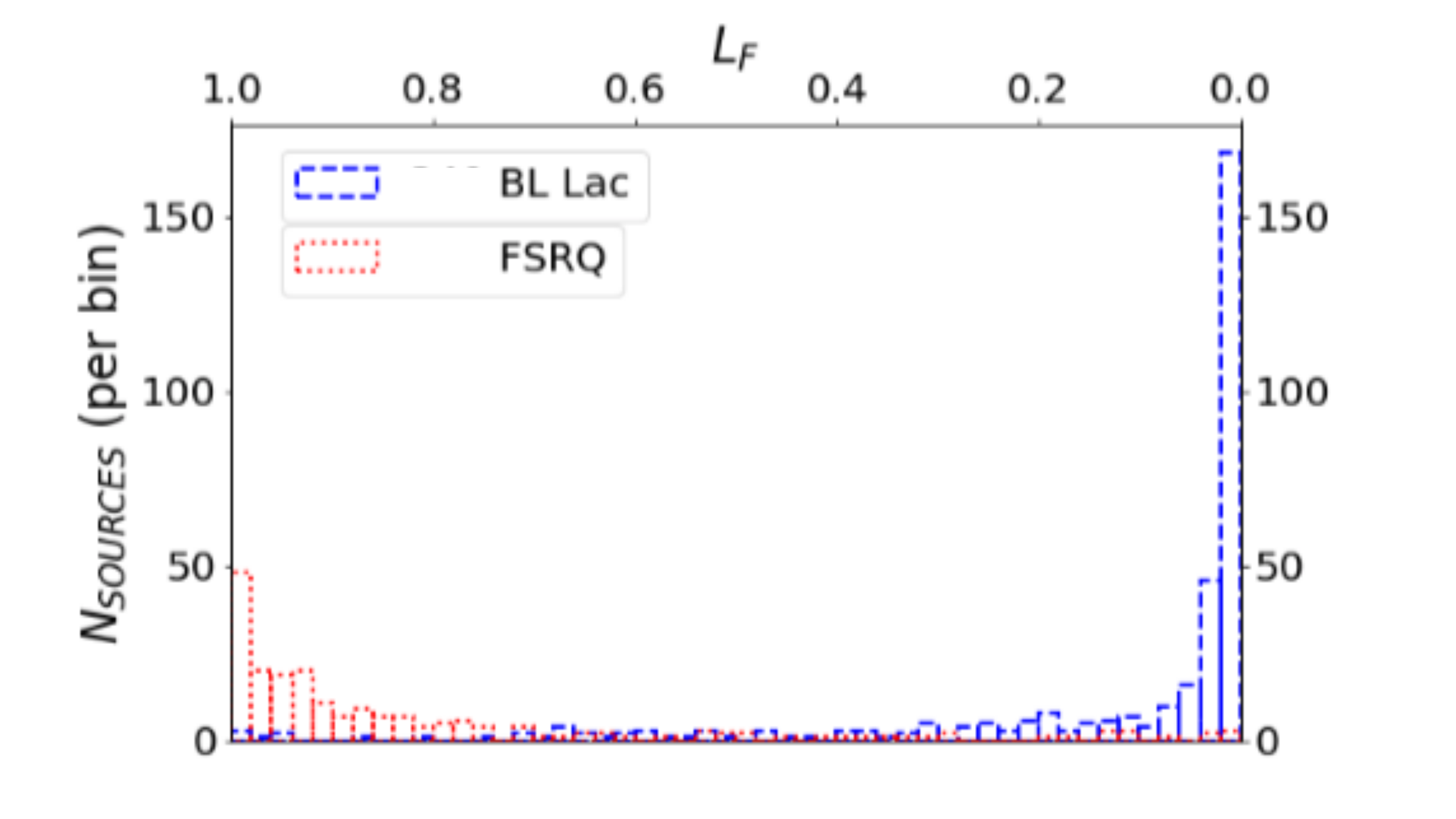} 
\caption{ Top: Histogram of the likelihood BL Lac (blue bar) and FSRQ
(red bar) of the UGS$_{BCU}$ sample. Two opposite and separate peaks differentiate the two classes of blazar.\label{ucsh}} 
\end{figure}

\section{Validation}
\label{<3>}
Since optical observations represent the most rigorous and reliable test for the classification of uncertain $\gamma$-ray sources, to validate our ANN analysis, we compared the results with the source classes inferred by \citet{fra} optical spectroscopic observations. In that study, the authors produced optical spectra for 130 4FGL BCUs and 50 unassociated $\gamma$-ray sources. 
The majority of the observed sources are found to be BL Lacs since they show a featureless optical spectrum. Our ANN predictions show an excellent agreement with the optical results.  
Only 9 predictions out of 180 observed sources misclassify the targets. 
The misclassification relates to particularly weak and unusually soft BL Lac sources
erroneously classified as FSRQ by the ANN algorithm.
For completeness, we also considered the 103 3FGL sources that remained uncertain in \citet{3zoo}. In 4FGL catalog 78 of those sources are classified as blazar, but 25 still remained as BCUs. Applying our ANN algorithm we classified 13 of them, obtaining 11 BL Lac and 2 FSRQ, while 12 sources remained BCUs. 
The optical spectra in \citet{fra} confirm the ANN classifications also for the latter sources showing that our ANN algorithm provides a reliable method to classify uncertain or unidentified gamma-ray sources also without optical observations.

\section{4FGLzoo}
\label{<4>}
Improving the results of the previous studies \citep{bcu, zhu, milos} with our UGS$_{BCU}$ analysis we predict a new classification (for the 4FGL $\gamma$-ray population 4FGLzoo)  with 1095 BL Lacs candidates and 570 FSRQs candidates while the fraction of the 4FGL uncertain sources decreases from 54$\%$ to 17$\%$. 
 In Table~\ref{ugs} we give a partial list of classified UGS$_{BCU}$  sources while the full list is available in the electronic format attached to this paper. 
It is also interesting to see, as shown in table~\ref{tabzoo}, how the relative contribution of the two blazar classes remains almost the same in both the eight years 4FGL catalog and in the four years 3FGL. 
By decreasing the uncertainty and increasing the sources class prediction the ANN  could be useful to the gamma-ray science community as a valid discriminant of best targets for future follow-up multi-wavelength observations. The full list of classified 4FGL BCUs  sources, analyzed in \citet{bcu}, is available in electronic format at {\url{https://cdsarc.unistra.fr/viz-bin/cat/J/MNRAS/493/1926}}

\begin{table}
\caption{Example of our classification for 10 4FGL UGS$_{BCU}$ sources. The full list is available in electronic format. Columns: 4FGL name, Galactic latitude,
Galactic longitude, L$_{B}$ , lower value of error interval L$_{B}$ low , upper value of error interval L$_{B}$ up \label{ugs}}
\label{table1}
\begin{center}
\begin{tabular}{lccccccccc}
\hline 
\hline
4FGL name 	&	b(deg)	& l(deg) & L$_{B}$& L$_{B}$
low & L$_{B}$ up \\
\hline		
4FGLJ0000.3-7355	&	-42.73	&	307.709	&	0.882	&	0.844	&	0.92 \\
4FGLJ0003.3+2511	&	-36.411	&	109.382	&	0.947	&	0.93	&	0.963 \\
4FGLJ0004.0+5715	&	-5.023	&	116.526	&	0.097	&	0.061	&	0.137 \\
4FGLJ0004.4-4001	&	-73.845	&	336.991	&	0.627	&	0.578	&	0.679	\\
4FGLJ0006.6+4618	&	-15.87	&	114.92	&	0.39	&	0.335	&	0.442	\\
4FGLJ0008.4+6926	&	6.886	&	119.148	&	0.335	&	0.207	&	0.466	\\
4FGLJ0008.9+2509	&	-36.725	&	110.915	&	0.016	&	0.002	&	0.039	\\
4FGLJ0009.1-5012	&	-65.545	&	319.395	&	0.236	&	0.171	&	0.297	\\
4FGLJ0009.2+1745	&	-43.965	&	108.909	&	0.23	&	0.15	&	0.312	\\
4FGLJ0009.7-1418	&	-73.912	&	83.706	&	0.37	&	0.285	&	0.454	\\

\hline
\end{tabular} \\
\end{center}
\end{table}

\begin{figure}
\begin{center}
\includegraphics[width=1.\textwidth]{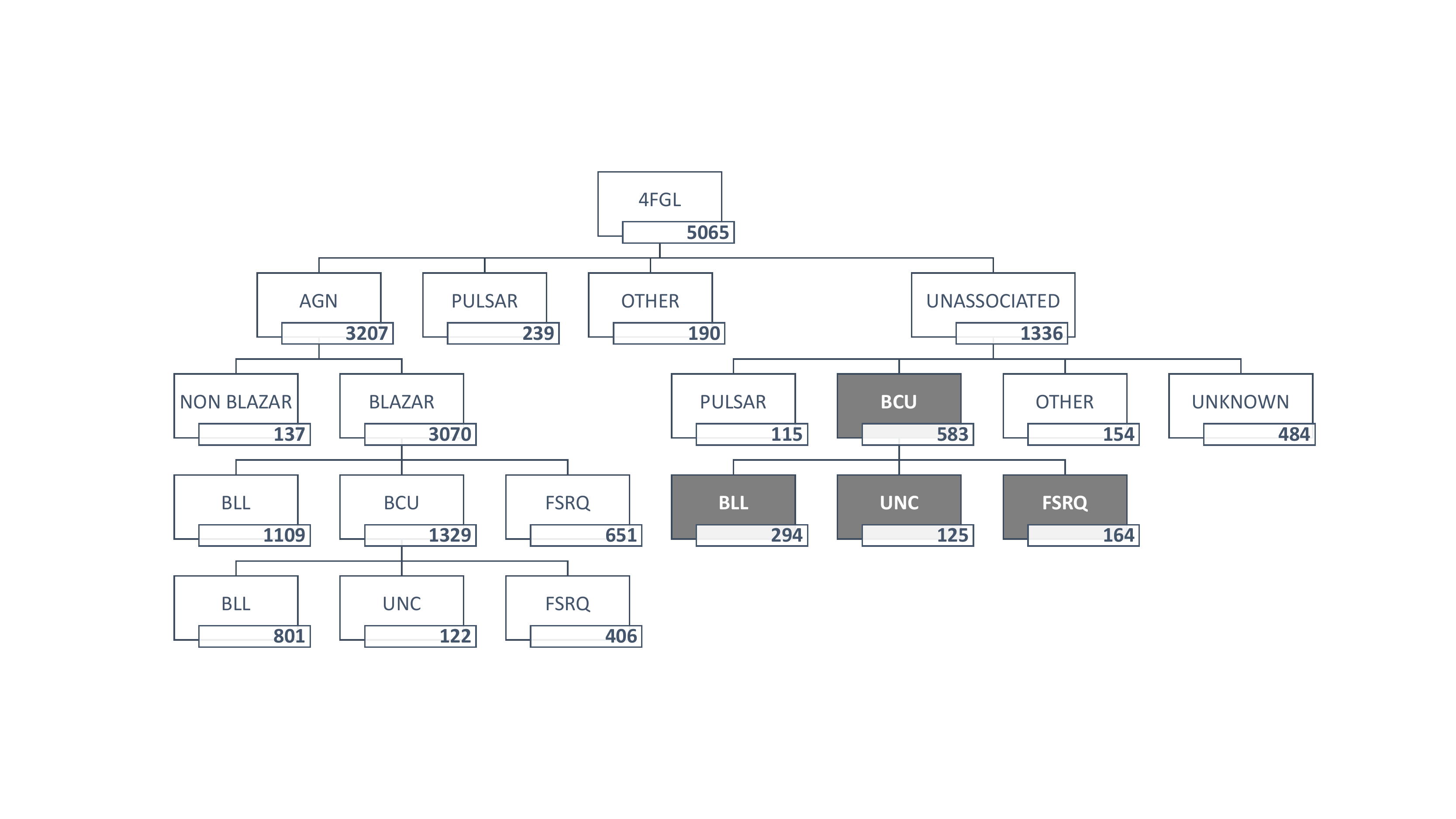}
\caption{{\bf{The 4FGLzoo. The ANN results of this study are highlighted as cells with a gray background.}} \label{zoo}}
\end{center}
\end{figure}

\begin{table}
\caption{The classification of the blazar
source classes in 4FGLzoo against the 3FGL
zoo after ANN analysis.\label{tabzoo}}
\begin{center}
\begin{tabular}{lcccccc}
\hline 
\hline
Class	&	3FGLzoo	& 4FGLzoo  \\
\hline		
Blazar	&	2279 	&	3672		\\
- BL Lac	& 1276 (56$\%$)		&	2204 (60$\%$)			\\
- FSRQ	&	823 (36$\%$)	& 1221 (33$\%$) \\
- BCU 	&	180 (8$\%$) 	&	247  (7$\%$)\\
\hline
\end{tabular} 
\end{center}
\end{table}

\section*{Acknowledgments}
\label{<5>}
Support for science analysis during the operation phase is gratefully acknowledged from the Fermi-LAT Collaboration for making the LAT results available in such a useful form. The authors gratefully thank Francesco Massaro  Dept of Physics, Università degli Studi, Torino, Italy and his collaborators for sharing optical spectroscopic material and Kerui Zhu, Dept.of Physics, Yunnan Normal University, ROC for sharing ML UGS results. The authors would like the anonymous referee for discussion and suggestions leading to the improvement of this work. \\


\end{document}